# Quantum gravity and information theories linked by the physical properties of the bit

Antonio Alfonso-Faus

Departamento de Aerotécnia, E.U.I.T. Aeronáutica

Plaza Cardenal Cisneros 3, 28040 Madrid, Spain

15 May 2011, aalfonsofaus@yahoo.es

**Abstract.-** Quantum gravity, and quantum cosmology, is not yet a complete nor consistent theory. One of the reasons for this is that the identification of the quantum of gravity is still very elusive. Here we show that the quantum of gravity is the minimum quantum of energy in nature, and we identify its physical properties. On the other hand, information theory has at its base the unit of information, the bit, and relies upon this entity that has no clear or definitely identified physical properties either. Here we prove that the bit has the absolute minimum of energy, and that it is the quantum of gravitational potential energy. Therefore the physics of the quantum of gravity and the physics of the quantum of information are parallel. We identify the entropy of 1 bit with the Boltzmann constant k, 61 orders of magnitude below the Planck scale as of today. It is the unit of entropy. These findings move forward the state of the art of two very important fields: quantum gravity and information theories. A new insight is also given for information entropy. We obtain these results thanks to the combination of two important contributions to science from the past: Weinberg relation and the Bekenstein maximum information limit.

Energy in nature is a well developed concept that has a dual property: it may be concentrated into a particle and/or manifested in a wave. A mass m has total relativistic energy mc² and as a particle it has a size r = ℏ/mc, its Compton wavelength (here c is the speed of light, ℏ is Planck´s constant and G the gravitational constant). As a wave it has an expression for its quantum of energy given by ℏc/λ. The self gravitational potential energy $E_g$ of a particle of mass m is given by

$$E_g = Gm^2/r \qquad (1)$$

Being the gravitational force extremely weak, compared to the electromagnetic and nuclear forces, one expects that its quantum must have an extremely small mass (energy) and momentum. The minimum possible quantum of energy, and momentum, that may exists in nature must have a maximum wavelength λ, that clearly is the size of the universe R (the maximum rage of any force). Then the minimum quantum of energy in the universe is therefore

$$E_m = ℏc/R \qquad (2)$$

We know that the energy of the gravitational field (and naturally its quantum) cannot be localized [1]. The only way to achieve this is to consider the extension of the quantum of gravity as equal to the size of the universe R: this gives a position uncertainty for this quantum of the order of R (it must be an unlocalized quantum). Then, the energy and momentum of the quantum of gravity must be

$$E_m = ℏc/R \qquad \text{and} \quad mc = ℏ/R \qquad (3)$$

The energy of this quantum is $mc^2$ and equating to $E_m$ in (2) we get for the equivalent mass of the quantum of gravity $m_g$

$$m_g = \hbar/cR \approx 10^{-66} \text{ grams} \qquad (4)$$

From the self gravitational potential energy (1) of a quantum particle of mass m and size $\hbar/mc = r$ we get now

$$E_g = Gm^2/r = Gm^2/(\hbar/mc) = Gm^3c/\hbar \qquad (5)$$

Weinberg´s relation [2] gives the mass m of any elementary particle in terms of quantum constants and the cosmological parameter R (or the equivalent expression with the Hubble parameter c/H) as

$$m^3 \approx \hbar^2/GR \qquad (6)$$

Eliminating m between (5) and (6) we get for the self gravitational potential energy of any elementary particle (having any mass m):

$$E_g = \hbar c/R \qquad (7)$$

Comparing (7) with (3), we have proved that the quantum of gravity is equal to the self gravitational potential energy of any elementary-particle of mass m, and is independent of m. Then it is a universal quantum, and a special elementary particle in the Weinberg sense. Its physical properties, energy, momentum and equivalent mass are given in (3) and (4).

Since the quantum of gravity has the minimum possible energy in nature, the number of equivalent gravity quanta in any mass M, with energy $E = Mc^2$, is

$$Mc^2/(\hbar c/R) = ER/\hbar c \qquad (8)$$

Multiplied by the constant $2\pi k$, k the Boltzmann constant, this is the maximum amount of information, the maximum number of bits given by the Bekenstein limit [3]. Then we may parallel the bit with the absolute minimum of energy, the quantum of gravity with physical properties (3) and (4).

For a black hole of mass M its entropy S is given by Hawking [4] as

$$S = 4\pi \, k/\hbar c \, GM^2 \qquad (9)$$

On the other hand, taking the universe as a black hole [5] from (9) and the black hole relation $2GM/c^2 = R$ its entropy is

$$S = 2\pi k \, Mc^2 R/\hbar c \qquad (10)$$

which is the Bekenstein[2] limit given in (8). It is clear now that 1 bit of information with entropy $S_1$ (and energy $\hbar c/R$) has the value

$$S_1 = 2\pi k \qquad (11)$$

Hence the Boltzmann constant k (times $2\pi$) is the unit of entropy, the bit, at the gravity quantum scale with mass

$$m_g = \hbar/cR \approx 10^{-66} \text{ grams} \qquad (12)$$

that clearly is not the Planck´s scale, as usually assumed. In fact the quantum of Planck, Planck´s units, has an entropy $S_p$ that we deduce here as

$$S_p \approx 2\pi k \, 10^{61} \qquad (13)$$

and for the universe $S_u$ has the well known value

$$S_u \approx 2\pi k \, 10^{122} \qquad (14)$$

The quantization of a black hole [6] gives the entropy $S_{bh}$ in terms of the number of nodes n, in a general state, as

$$S_{bh} = 2\pi k \, n \qquad (15)$$

and considering again the universe as a black hole [5] we get from (14) and (15) the number of nodes in our universe, at present, as

$$n = 10^{122} \qquad (16)$$

Hence, in a sense, the universe may be thought to be a black hole in the excited state to day given by (16). It is clear that for the ground state, n = 1, we get the entropy of 1 bit $2\pi k$.

**Conclusions**

We have found here an expression for the minimum quantum of energy in nature. We have identified it to the quantum of gravity and to the quantum of information, the bit. These findings have consequences for improving the state of the art of quantum gravity and information theories. And as far as cosmology is concerned, we find that our universe resembles the excited state of a quantum black hole.